%% file: main.tex
\documentclass[%
 reprint,
 superscriptaddress,
nofootinbib,
 amsmath,amssymb,
 aps,
 floatfix,
10pt]{revtex4-2}
\usepackage[utf8]{inputenc}
\usepackage{graphicx}
\usepackage{subfig}
\usepackage{dcolumn}
\usepackage{bm}
\usepackage{booktabs}
\usepackage{pgf}
\usepackage{pgfplots}
\pgfplotsset{compat=newest}
\usepackage{mathtools,slashed}
\usepackage{hyperref}
\usepackage{xfrac}
\usepackage{xcolor}
\usepackage{gensymb}
\usepackage{subfig}
\usepackage{caption}
\usepackage[normalem]{ulem}
\usepackage{fancyhdr}
 \fancypagestyle{plain}{%
 	\fancyhf{}
 	\fancyfoot[C]{\thepage}
 }

\begin{document}

\preprint{APS/123-QED}

\title{Azimuthal asymmetry in exclusive quasi-elastic neutrino-nucleus interactions}
\author{Marco Vanderpoorten}
\email{marco.vanderpoorten@ugent.be}
\affiliation{Department of Physics and Astronomy, Ghent University, Proeftuinstraat 86, B-9000 Gent, Belgium}
\author{Ashish Kumar Jha}
\affiliation{Department of Physics and Astronomy, Ghent University, Proeftuinstraat 86, B-9000 Gent, Belgium}
\author{Mathias El Baz}
\affiliation{Département de Physique Nucléaire et Corpusculaire, Université de Genève, Geneva, Switzerland}
\author{Kajetan Niewczas}
\affiliation{École polytechnique, IN2P3-CNRS, Laboratoire Leprince-Ringuet, F-91120 Palaiseau, France}
\affiliation{Department of Physics and Astronomy, Ghent University, Proeftuinstraat 86, B-9000 Gent, Belgium}
\author{Federico Sánchez}
\affiliation{Département de Physique Nucléaire et Corpusculaire, Université de Genève, Geneva, Switzerland}
\author{Natalie Jachowicz}
\email{natalie.jachowicz@ugent.be}
\affiliation{Department of Physics and Astronomy, Ghent University, Proeftuinstraat 86, B-9000 Gent, Belgium}

\date{\today}

\begin{abstract}
In neutrino oscillation experiments, exclusive measurements of neutrino–nucleus interactions play a critical role, by providing the theoretical and experimental input needed for a reliable estimation of the neutrino energy. In this paper, we derive the general form of the azimuthal angle distribution for quasi-elastic scattering, focusing on a dependency that has been routinely overlooked. We demonstrate that the outgoing nucleon exhibits a preference for emission outside the lepton scattering plane, with an  asymmetric azimuthal  distribution. In the context of neutrino–nucleus scattering, we argue that this asymmetry is caused by parity violation in the weak interaction. Furthermore, we show that in cross section calculations the asymmetry is sensitive to nuclear modeling choices and to the shell structure of the initial nucleus, thus providing a novel source of information for energy reconstruction in neutrino experiments. We study the experimental feasibility of observing this effect by applying a realistic momentum detection threshold and an intranuclear cascade. We estimate that the asymmetry is observable with $\mathcal{O}(10^4)$ events at the $99\%$ confidence level for neutrino interactions on $^{12}$C, suggesting that the effect is within reach of the current generation of neutrino detectors.

\end{abstract}
\maketitle
\thispagestyle{empty}

\input{Sections/Introduction}
\input{Sections/Formalism}
\input{Sections/Results}

\input{Sections/Experiment}

\input{Sections/Conclusion}

\begin{acknowledgments}
The authors acknowledge support from  Ghent University Special Research Fund and the Fund for Scientific Research Flanders (FWO-Flanders). We gratefully acknowledge the Swiss National Science Foundation (SNSF) for financial support under LEAD program grant number 200021E\_213196.  
\end{acknowledgments}

\appendix

\section{Hadron tensor}
\label{Apprespons}

Although this paper focuses on CCQE interactions, the following derivation is general for all interactions X($\nu_l, l^- x$) and X($\overline{\nu}_l, l^+ x$) where one lepton and one hadron is observed. Here, $X$ denotes the target nucleus, $l$ any lepton and $x$ any measured kinematically allowed particle. Note that the presence of only one outgoing particle does not necessarily mean that only this particle was emitted, but rather that only one particle was detected.

It is customary to denote the symmetric (unprimed) and antisymmetric hadronic response functions (primed) as follows:

\begin{align}
    W^{CC} &\equiv \mathcal{R}(W^{00}) = W^{00}, \nonumber \\
    W^{CL} &\equiv 2\mathcal{R}(W^{03}) = 2W^{03}_s, \nonumber\\
    W^{LL} &\equiv \mathcal{R}(W^{33}) = W^{33},\nonumber\\
    W^{T} &\equiv \mathcal{R}(W^{22}) + \mathcal{R}(W^{11}) = W^{22} + W^{11}, \nonumber\\
    W^{TT} &\equiv \mathcal{R}(W^{22}) - \mathcal{R}(W^{11}) = W^{22} - W^{11} ,\nonumber\\
    W^{TC} &\equiv 2\sqrt{2}\mathcal{R} (W^{01}) = 2\sqrt{2}W^{01}_s,\nonumber\\
    W^{TL} &\equiv 2\sqrt{2}\mathcal{R} (W^{31}) = 2\sqrt{2}W^{31}_s,\nonumber\\
    W^{\underline{T}T} &\equiv 2 \mathcal{R}(W^{12}) = 2 W^{12}_s ,\nonumber\\
    W^{\underline{T}C} &\equiv 2\sqrt{2} \mathcal{R}(W^{02}) = 2\sqrt{2} W^{02}_s,\nonumber\\
    W^{\underline{T}L} &\equiv 2\sqrt{2} \mathcal{R}(W^{32}) = 2\sqrt{2} W^{32}_s,\nonumber\\
    W^{T'} &\equiv -2 \mathcal{I}(W^{12}) = 2iW^{12}_a,\nonumber \\
    W^{TC'} &\equiv -2\sqrt{2} \mathcal{I}(W^{02}) = 2\sqrt{2}iW^{02}_a,\nonumber \\
    W^{TL'} &\equiv -2\sqrt{2} \mathcal{I}(W^{32}) = 2\sqrt{2}iW^{32}_a ,\nonumber\\
    W^{\underline{C}L'} &\equiv \mathcal{I}(W^{03}) = i W^{03}_a,\nonumber\\
    W^{\underline{T}C'} &\equiv 2\sqrt{2}\mathcal{I}(W^{01}) = -2\sqrt{2}i W^{01}_a,\nonumber\\
    W^{\underline{T}L'} &\equiv 2\sqrt{2}\mathcal{I}(W^{31}) = -2\sqrt{2}i W^{31}_a  .  
\end{align}

Using these definitions, it is straightforward to show that these response functions adopt the following form by calculating Eqs.~(\ref{respeq1}) to (\ref{respeq4}):\\

\newpage
$\underline{{(W^{\mu\nu})^{PC}_s}_{VV/AA}}$

\begin{align}
\label{resp1}
    W^{CC} &= W_1 + \frac{1}{\rho^2} W_2 + \frac{\nu^2}{\rho} W_3 + \frac{H^2}{\rho^2} W_4 + 2\frac{\nu}{\rho^{3/2}} W_5 \nonumber \\
&\quad+ 2 \frac{H \nu}{\rho^{3/2}} W_6 + 2\frac{H}{\rho^2} W_7, \nonumber\\
    W^{CL} &= 2\Big\{ \frac{\nu}{\rho^2} W_2 + \frac{\nu}{\rho}W_3 + \frac{\nu H^2}{\rho^2} W_4 + \frac{1}{\rho^{3/2}}(1 + \nu^2) W_5 \nonumber \\
&\quad+ \frac{H}{\rho^{3/2}}(1 + \nu^2) W_6 + \frac{2H\nu}{\rho^2} W_7 \Big\}, \nonumber\\
    W^{LL} &= - W_1 + \frac{\nu^2}{\rho^2} W_2 + \frac{1}{\rho} W_3 + \frac{\nu^2 H^2}{\rho^2} W_4 + 2\frac{\nu}{\rho^{3/2}} W_5\nonumber \\
&\quad + 2 \frac{H \nu}{\rho^{3/2}} W_6 + 2\frac{H \nu^2}{\rho^2}W_7, \nonumber\\
    W^T &= -2 W_1 + \eta_T^2 W_4, \nonumber\\
    W^{TT} &= - \eta_T^2 W_4 \cos2\phi_N, \nonumber\\
    W^{TC} &= 2\sqrt{2}\cos\phi_N \Big\{ \frac{H \eta_T}{\rho} W_4 + \frac{\nu \eta_T}{\sqrt{\rho}} W_6 + \frac{\eta_T}{\rho} W_7 \Big\}, \nonumber\\
    W^{TL} &= 2\sqrt{2}\cos\phi_N \Big\{ \frac{\nu \eta_T H}{\rho} W_4 + \frac{\eta_T}{\sqrt{\rho}}W_6 + \frac{\nu \eta_T}{\rho} W_7 \Big\}, \nonumber\\
    W^{\underline{T}T} &= \eta_T^2 W_4 \sin2\phi_N ,\nonumber\\
    W^{\underline{T}C} &= 2\sqrt{2}\sin\phi_N \Big\{ \frac{H \eta_T}{\rho} W_4 + \frac{\nu \eta_T}{\sqrt{\rho}} W_6 + \frac{\eta_T}{\rho} W_7 \Big\}, \nonumber\\
    W^{\underline{T}L} &= 2\sqrt{2}\sin\phi_N \Big\{ \frac{\nu \eta_T H}{\rho} W_4 + \frac{\eta_T}{\sqrt{\rho}} W_6 + \frac{\nu \eta_T}{\rho}W_7\Big\},
\end{align}

$\underline{{(W^{\mu\nu})^{PV}_s}_{VA}}$
 
\begin{align}
\label{resp2}
    W^{CC} &= 0 ,\nonumber\\
    W^{CL} &= 0,\nonumber\\
    W^{LL} &= 0 ,\nonumber\\
    W^T &= 0, \nonumber\\
        W^{TT} &= - \frac{\eta_T^2}{\sqrt{\rho}} W_{10} \sin2\phi_N ,\nonumber\\
    W^{TC} &= 2\sqrt{2}\sin\phi_N \Big\{ \frac{\nu \eta_T}{\rho} W_8 + \frac{\eta_T}{\rho^{3/2}} W_9 + \frac{H \eta_T}{\rho^{3/2}} W_{10} \Big\} ,\nonumber\\
    W^{TL} &= 2\sqrt{2}\sin\phi_N \Big\{ \frac{\eta_T}{\rho} W_8 + \frac{\nu \eta_T}{\rho^{3/2}}W_9 + \frac{\nu H \eta_T}{\rho^{3/2}} W_{10} \Big\},\nonumber \\
    W^{\underline{T}T} &= -2 \frac{\eta_T^2}{\sqrt{\rho}} W_{10} \cos2\phi_N, \nonumber\\
    W^{\underline{T}C} &= 2\sqrt{2}\cos\phi_N \Big\{ \frac{\nu \eta_T}{\rho} W_8 + \frac{\eta_T}{\rho^{3/2}} W_9 + \frac{H \eta_T}{\rho^{3/2}} W_{10} \Big\},\nonumber \\
    W^{\underline{T}L} &= 2\sqrt{2}\cos\phi_N \Big\{ \frac{\eta_T}{\rho} W_8 + \frac{\nu \eta_T}{\rho^{3/2}} W_9 + \frac{\nu H \eta_T}{\rho^{3/2}}W_{10}\Big\},
\end{align}\\

$\underline{{(W^{\mu\nu})^{PV}_a}_{VV/AA}}$
 
\begin{align}
\label{resp3}
    W^{T'} &= 0 ,\nonumber\\
    W^{TC'} &= -2\sqrt{2}\sin\phi_N \Big\{\frac{\nu \eta_T}{\sqrt{\rho}} W_{12} + \frac{\eta_T}{\rho} W_{13} \Big\},\nonumber \\
    W^{TL'} &=  - 2\sqrt{2}\sin\phi_N \Big\{\frac{\eta_T}{\sqrt{\rho}} W_{12} + \frac{\nu \eta_T}{\rho} W_{13} \Big\},\nonumber\\
    W^{\underline{C}L'} &=  \frac{1}{\sqrt{\rho}}
    W_{11}+ \frac{H}{\sqrt{\rho}} W_{12} ,\nonumber\\
    W^{\underline{T}C'} &= 2\sqrt{2}\cos\phi_N \Big\{ \frac{\nu \eta_T}{\sqrt{\rho}} W_{12} + \frac{\eta_T}{\rho} W_{13} \Big\}, \nonumber\\
    W^{\underline{T}L'} &=  2\sqrt{2}\cos\phi_N \Big\{ \frac{\eta_T}{\sqrt{\rho}} W_{12} + \frac{\nu \eta_T}{\rho} W_{13} \Big\},
\end{align}


$\underline{{(W^{\mu\nu})^{PC}_a}_{VA}}$
 
\begin{align}
\label{resp4}
    W^{T'} &=  - \frac{2}{\sqrt{\rho}} W_{14} + \frac{2H}{\sqrt{\rho}} W_{16} + \frac{2 \eta_T^2}{\sqrt{\rho}} W_{19},\nonumber \\
    W^{TC'} &=  -2\sqrt{2}\cos\phi_N \Big\{- \frac{\nu \eta_T}{\rho} W_{15} - \frac{\eta_T}{\sqrt{\rho}} W_{16} - \frac{\nu \eta_T}{\rho} W_{17} \nonumber \\
&\quad- \frac{\eta_T}{\rho^{3/2}} W_{18} - \frac{H \eta_T}{\rho^{3/2}} W_{19} \Big\},\nonumber \\
    W^{TL'} &= - 2\sqrt{2}\cos\phi_N \Big\{- \frac{\eta_T}{\rho} W_{15} - \frac{\nu \eta_T}{\sqrt{\rho}} W_{16} - \frac{\eta_T}{\rho} W_{17} \nonumber \\
&\quad- \frac{\nu \eta_T}{\rho^{3/2}} W_{18} - \frac{\nu H \eta_T}{\rho^{3/2}} W_{19} \Big\},\nonumber\\
    W^{\underline{C}L'} &= 0,\nonumber\\
    W^{\underline{T}C'} &=  2\sqrt{2}\sin\phi_N \Big\{\frac{\nu \eta_T}{\rho} W_{15} + \frac{\eta_T}{\sqrt{\rho}} W_{16} + \frac{\nu \eta_T}{\rho} W_{17}\nonumber \\
&\quad + \frac{\eta_T}{\rho^{3/2}} W_{18} + \frac{H \eta_T}{\rho^{3/2}} W_{19} \Big\} ,\nonumber\\
    W^{\underline{T}L'} &= 2\sqrt{2} \sin\phi_N \Big\{ \frac{\eta_T}{\rho} W_{15} + \frac{\nu \eta_T}{\sqrt{\rho}} W_{16} + \frac{\eta_T}{\rho} W_{17} \nonumber \\
&\quad+ \frac{\nu \eta_T}{\rho^{3/2}}W_{18} + \frac{\nu H \eta_T}{\rho^{3/2}} W_{19} \Big\},
\end{align}
where the following definitions were used:
\begin{align}
\nu &\equiv \frac{\omega}{q},\\
\rho&\equiv \frac{|Q^2|}{q^2} = 1 - \nu^2 ,\\
\eta_T &\equiv \frac{p_N}{M_N} \sin\theta_N ,\\
H &\equiv \frac{1}{M_N} \left[E_N - \nu p_N \cos\theta_N\right].
\end{align}

Not all of these terms are equally important in the calculation of the cross section. The lepton responses corresponding to the underlined hadronic responses ($W^{\underline{C}L}$, $W^{\underline{T}C}$, ...) are antisymmetric. They vanish when the incoming particles are longitudinally polarized or if the polarization of the incoming and final particles is not specified~\cite{TheMorenoPaper,donneleyraskin,DonRas2}. As neutrinos are longitudinally polarized particles, these underlined hadronic responses vanish in neutrino scattering. The azimuthal angle dependence of the relevant response functions can then be summarized as in Table~\ref{responssummary}. \\

\begin{table}[h!]
	\centering
		\caption{Summary of the azimuthal angle dependence of the relevant response functions in the calculation of the cross sections. The sine-dependencies here induce parity violating effects.}
	\renewcommand{\arraystretch}{1.3} 
	\resizebox{0.27\textwidth}{!}{
		\begin{tabular}{l|l|l|l}
			& VV & AA & VA  \\ 
			\hline
			$W^{\!CC}$ & const. & const.  &  0 \\
			$W^{\!CL}$ & const.  & const.  &  0 \\
			$W^{\!LL}$ & const.  & const.  &   0   \\
			$W^{\!T}$ & const. & const. & 0 \\
			$W^{\!TT}$ & $\cos2\phi_N$ & $\cos2\phi_N$ & $\sin2\phi_N$ \\
			$W^{\!TC}$ & $\cos\phi_N$ & $\cos\phi_N$ & $\sin\phi_N$ \\
			$W^{\!TL}$ & $\cos\phi_N$ & $\cos\phi_N$ & $\sin\phi_N$ \\
			$W^{\!T'}$ & 0 & 0 & const. \\
			$W^{\!TC'}$ & $\sin\phi_N$ & $\sin\phi_N$ & $\cos\phi_N$ \\
			$W^{\!TL'}$ & $\sin\phi_N$ & $\sin\phi_N$ & $\cos\phi_N$ \\
		\end{tabular}
	}
	\label{responssummary}
\end{table}

\section{Azimuthal angle distribution in PWIA}
\label{PWIA}

Considering the definitions of the hadron response tensor and nuclear current from Eqs.~(\ref{hadtens}) and (\ref{currentdef}), it is clear that an expression for the incoming and outgoing waves is necessary to calculate these quantities. In the relativistic plane wave impulse approximation RPWIA, one considers these waves to be plane waves, which can be represented by free Dirac spinors. In the non-relativistic PWIA however, the antiparticle spinors are projected out. The following derivation is restricted to the positive energy solutions of the Dirac equation, the extension to RPWIA is trivial. The hadron response tensor then becomes:
\begin{align}
    4m_N^2 W^{\mu \nu} = \text{Tr} \Bigg[(\slashed{P}_N+m_N)\hat{J}^\mu(\overline{\slashed{P}} + m_N)\overline{\hat{J}}^{\lower1.0ex\hbox{$\scriptstyle \nu$}}
\Bigg],
    \label{start}
\end{align}
where in general $\overline{P}^\mu = \gamma^0P^\mu\gamma^0$. $\hat{J}^\mu$ is the nuclear current operator as in Eq.~(\ref{nucloperator}). $P_N^\mu$ and $\overline{P}^\mu$ are the four momenta of the nucleon before and after the interaction respectively. This leads to:

\begin{align}
    4m_N^2 W^{\mu\nu} &= \text{Tr} \Bigg[ 
(\slashed{P}_N + m_N) 
\Big( F_1(|Q^2|) \gamma^\mu \nonumber \\ &+ \frac{i F_2(|Q^2|)}{2 m_N} \sigma^{\mu\alpha} Q_\alpha  + G_A(|Q^2|) \gamma^\mu \gamma^5 \nonumber \\ &+ \frac{G_P(|Q^2|)}{2 m_N} \gamma^5 Q^\mu \Big) 
(\overline{\slashed{P}} + m_N) \nonumber\\
&\Big( F_1(|Q^2|) \gamma^\nu - \frac{i F_2(|Q^2|)}{2 m_N} \sigma^{\nu\beta} Q_\beta \nonumber \\ &+ G_A(|Q^2|) \gamma^\nu \gamma^5 - \frac{G_P(|Q^2|)}{2 m_N} \gamma^5 Q^\nu \Big)
\Bigg].
\end{align}

The distinct contributions to the hadron tensor can now be separated into VV-, AA- and VA-parts to compare the results with the derivation using the invariant functions:

 \begin{widetext}
\begin{align}
    4m_N^2 W_{VV}^{\mu\nu} =& \text{Tr}\Bigg[(\slashed{P}_N + m_N)\Big(F_1(|Q^2|) \gamma^\mu + \frac{iF_2(|Q^2|)}{2m_N}\sigma^{\mu\alpha}Q_\alpha\Big)(\overline{\slashed{P}} + m_N)\Big(F_1(|Q^2|) \gamma^\nu - \frac{iF_2(|Q^2|)}{2m_N}\sigma^{\nu\beta}Q_\beta\Big)\Bigg],\\
    4m_N^2 W_{AA}^{\mu\nu} =& \text{Tr}\Bigg[(\slashed{P}_N + m_N)\Big(G_A(|Q^2|)\gamma^\mu\gamma^5+\frac{G_P(|Q^2|)}{2m_N}\gamma^5Q^\mu\Big)(\overline{\slashed{P}} + m_N)\Big(G_A(|Q^2|)\gamma^\nu\gamma^5-\frac{G_P(|Q^2|)}{2m_N}\gamma^5Q^\nu\Big)\Bigg],\\
    4m_N^2W_{VA}^{\mu\nu}=&\text{Tr}\Bigg[(\slashed{P}_N + m_N)\Big(F_1(|Q^2|) \gamma^\mu + \frac{iF_2(|Q^2|)}{2m_N}\sigma^{\mu\alpha}Q_\alpha\Big)(\overline{\slashed{P}} + m_N)\Big(G_A(|Q^2|)\gamma^\nu\gamma^5-\frac{G_P(|Q^2|)}{2m_N}\gamma^5Q^\nu\Big)\Bigg] +\nonumber\\&\text{Tr}\Bigg[(\slashed{P}_N + m_N)\Big(G_A(|Q^2|)\gamma^\mu\gamma^5+\frac{G_P(|Q^2|)}{2m_N}\gamma^5Q^\mu\Big)(\overline{\slashed{P}} + m_N)\Big(F_1(|Q^2|) \gamma^\nu - \frac{iF_2(|Q^2|)}{2m_N}\sigma^{\nu\beta}Q_\beta\Big)\Bigg].
\end{align}

Using the necessary trace identities, this ultimately leads to (For a more detailed derivation, see eg.~\cite{patinothesis}):

\begin{align}
    m_N^2 W^{\mu\nu}_{VV} =& F_1(|Q^2|)^2 \left( \overline{P}^\mu P_N^\nu + \overline{P}^\nu P_N^\mu + \frac{\overline{Q}^2}{2} g^{\mu\nu} \right) 
 + F_1(|Q^2|) F_2(|Q^2|) \big( Q \cdot \overline{Q} g^{\mu\nu} \nonumber- \frac{Q^\mu \overline{Q}^\nu + Q^\nu \overline{Q}^\mu}{2} \big)  \nonumber\\&+ \frac{F_2(|Q^2|)^2}{4m_N^2} \Bigg[ P_N \cdot Q \big( \overline{P}^\mu Q^\nu + \overline{P}^\nu Q^\mu \big) 
+ \overline{P} \cdot Q \big( P_N^\mu Q^\nu + P_N^\nu Q^\mu \big) 
- Q^2 \left( P_N^\mu \overline{P}^\nu + P_N^\nu \overline{P}^\mu \right) \nonumber \\
&
- \left( 2m_N^2 - \frac{\overline{Q}^2}{2} \right) Q^\mu Q^\nu 
+ g^{\mu\nu} \big( 2m_N^2 Q^2 - \frac{Q^2 \overline{Q}^2}{2}- 2 (P_N \cdot Q)(\overline{P} \cdot Q) \big) \Bigg], \\
m_N^2 W^{\mu\nu}_{AA} =& G_A(|Q^2|)^2 \Bigg[ P_N^\mu \overline{P}^\nu + P_N^\nu \overline{P}^\mu 
- g^{\mu\nu} \big( 2m_N^2 - \frac{\overline{Q}^2}{2} \big) \Bigg] 
\nonumber \\ &- \frac{G_P(|Q^2|)^2 \overline{Q}^2}{8m_N^2} Q^\mu Q^\nu 
- \frac{G_A(|Q^2|) G_P(|Q^2|)}{2} \Bigg[ \overline{Q}^\mu Q^\nu + \overline{Q}^\nu Q^\mu \Bigg],\\
m_N^2 W^{\mu\nu}_{VA} =& i \Bigg[ G_A(|Q^2|) \epsilon^{\mu\nu\alpha\beta} 
\big( 2 F_1(|Q^2|) P_{N\alpha} \overline{P}_\beta - F_2(|Q^2|) (P_N + \overline{P})_\alpha Q_\beta \big) 
\nonumber \\ &+ \frac{G_P(|Q^2|) F_2(|Q^2|)}{4m_N^2} \big( -Q^\mu \epsilon^{\nu\alpha\beta\sigma} 
+ Q^\nu \epsilon^{\mu\alpha\beta\sigma} \big) P_{N\alpha} \overline{P}_\sigma Q_\beta \Bigg]. 
\end{align}
 
\end{widetext}

The vectors in these equations are defined as follows: 

\begin{align}
    P_N = (&E_N, p_N \sin\theta_N\cos\phi_N, \nonumber \\
&\sin\theta_N\sin\phi_N,p_N\cos\theta_N), \\
    \overline{P} = (&\overline{E}, p_N \sin\theta_N\cos\phi_N,\nonumber \\&\sin\theta_N\sin\phi_N,p_N\cos\theta_N-q),\\
    Q^\mu = (&\omega,0,0,q).
\end{align}

Note that in PWIA, the x- and y-components of the initial and outgoing nucleon's momenta are equal, as the interaction only happens in the z-direction and there is no influence of the nuclear potential on the outgoing nucleon. Using these vector definitions, the azimuthal angle distributions of the relevant hadronic response functions are given by:
\begin{align}
    W^{CC} &= W^{00} = W^{CC}_{VV} + W^{CC}_{AA}, \nonumber\\
    W^{CL}& = 2W_s^{03} = W^{CL}_{VV} + W^{CL}_{AA}, \nonumber\\
    W^{LL} &= W^{33} = W^{LL}_{VV} + W^{LL}_{AA}, \nonumber\\
    W^{T} &= W^{22} + W^{11} = W^{T}_{VV} + W^{T}_{AA}, \nonumber \\
    W^{TT} &= W^{22} - W^{11} \nonumber \\ &= W^{TT}_{VV}\cos2\phi_N + W^{TT}_{AA}\cos2\phi_N, \nonumber\\
    W^{TC} &= 2\sqrt{2}W_s^{01}\nonumber\\&=W^{TC}_{VV}\cos\phi_N + W^{TC}_{AA}\cos\phi_N, \nonumber \\
     W^{TL} &=2\sqrt{2}W_s^{31} \nonumber \\&= W^{TL}_{VV}\cos\phi_N + W^{TL}_{AA}\cos\phi_N,\nonumber \\
     W^{T'} &= 2iW_a^{12}=W^{T'}_{VA}, \nonumber\\
     W^{TC'} &= 2\sqrt{2}i W_a^{02}=W^{TC'}_{VA}\cos\phi_N, \nonumber\\
     W^{TL'} &= 2\sqrt{2}iW_a^{32}=W^{TL'}_{VA}\cos\phi_N,
\end{align} 
where the $W$s with a subscript represent all other functional structures besides the azimuthal angle dependence of the parts of the hadronic responses. Here, the $\sin\phi_N$ dependencies from Table~\ref{responssummary} for the general case disappear as these terms cancel out due to the equality between the x- and y-components of the initial and final nucleon momenta.

%
%

%
\bibliography{references}

\end{document}

%% file: Sections/Introduction.tex
\section{Introduction}
Neutrino oscillation experiments have increasingly entered what may be called the ‘exclusive era’. Among others, T2K~\cite{T2K:2018rnz,intro5}, NINJA~\cite{NINJA:2020gbg}, MINER$\nu$A~\cite{MINERvA:2023avz,intro2,intro3,intro4}, SBND~\cite{acciarri2015proposaldetectorshortbaselineneutrino}, ArgoNeuT~\cite{intro1} and MicroBooNE~\cite{MicroBooNE:2025ooi,microboone} detectors have proven the capability to detect the final-state lepton ($\mu^\pm, e^\pm$) in coincidence with other charged particles ($p^+, \pi^{\pm},...$).  The hadron kinematic variables, often referred to as  exclusive variables, provide additional information on the final-state particles, help constrain the neutrino energy reconstruction~\cite{modelbelangrijk,Douqa:2024pmy}, and even improve our understanding of the neutrino beam~\cite{Karpova:2025lvb}.

In the reconstruction of the neutrino energy, precise neutrino-nucleus interaction cross sections are pivotal. Typically, Monte Carlo event generators are tuned to match data from near detectors, and the results are extrapolated to the far detector. However, recent studies have shown that this procedure is insufficient in reducing the systematic uncertainties in the determination of physical parameters~\cite{Uncertain-Alg,Uncertain-BSM,Uncertain-SN}. A novel approach for studying the relationship between the detector’s observable energy deposits and the true energy of the interacting neutrino is the Precision Reaction Independent Spectrum Measurement (PRISM)~\cite{DUNE:2025lvs} system, which will be implemented in DUNE~\cite{DUNE:2016hlj,DUNE:2015lol,DUNE:2016evb,DUNE:2016rla}, or the Intermediate Water Cherenkov Detector (IWCD)~\cite{Scott:2016kdg} that will be part of the Hyper-K~\cite{Hyper-Kamiokande:2025fci} experiment. By positioning the near detector at different off-axis angles relative to the beam, PRISM and IWCD enable measurements of varied neutrino spectra~\cite{prism,Zhu:2023nsv}. While experimental strategies such as PRISM and IWCD can help constrain the relation between detector observables and neutrino energy, theoretical modeling remains indispensable for describing and understanding the complex nuclear dynamics involved in neutrino-nucleus interactions.\\

To provide the theoretical and experimental input needed for reliable energy reconstruction, exclusive measurements of neutrino–nucleus interactions play a central role. In such measurements, the outgoing charged lepton is reconstructed in coincidence with at least one other particle from the interaction. A particularly important example is the Charged-Current Quasi-Elastic (CCQE) process. These are interactions of the form: $\nu_l + ^A_ZX \to l^- + ^{A-1}_ZX + p^+$ or $\overline{\nu}_l + ^A_ZX \to l^+ + ^{A-1}_{Z-1}Y + n^0$, with $l$ any charged lepton (e, $\mu$ or $\tau$). These interactions account for a significant part of the total cross section in most current (T2K~\cite{T2Kdesign}, MicroBooNE~\cite{microboonedesign}, MINER$\nu$A~\cite{MINERvA:2013zvz}) and future (Hyper-Kamiokande, DUNE, ESS$\nu$SB~\cite{essnusb}) accelerator-based neutrino detectors. In such processes, five independent variables can be considered: two lepton kinematic variables and three hadron kinematic variables. The former are detected in inclusive measurements, while the latter  constitute the hadronic exclusive observables. These are the hadron kinetic energy, polar and azimuthal angle. In this work, we investigate how hadron observables can improve our understanding of neutrino-nucleus interactions. In particular, we focus on the final-state hadron's azimuthal angle and how the dependence of the cross section on this quantity varies with nuclear shell structure differences and modeling choices. 
As the carbon-to-oxygen cross section ratio is vital for water-based experiments (e.g., Hyper-K, Super-K~\cite{ratioOC}), studying the azimuthal angle distribution can significantly enhance carbon-to-oxygen migration models.

The paper is organized as follows. In Sect.~\ref{twee}, the kinematics of quasi-elastic neutrino–nucleus scattering is reviewed and the general dependence of the cross section on the azimuthal angle of the outgoing nucleon is analyzed. In Sect.~\ref{drie}, the nuclear models employed in this work are discussed, and their respective azimuthal angle dependencies are examined. The results of the present study and their implications for the validation of nuclear models are presented in Sect.~\ref{vier}. The feasibility of observing the predicted effects in current long-baseline neutrino experiments is assessed in Sect.~\ref{vierb}. Our conclusions are summarized in the final Sect.~\ref{vijf}.

%% file: Sections/Formalism.tex
\section{Theory}
\label{twee}
\subsection{Kinematics and cross section}

\begin{figure}[htbp]
  \centering
  \includegraphics[width=0.48\textwidth]{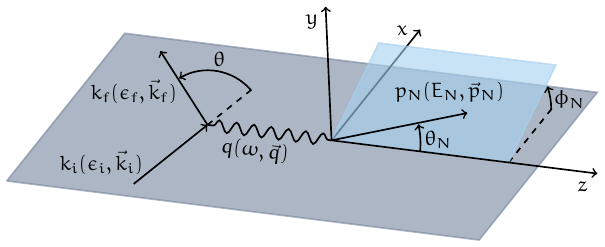} 
  \caption{Overview of the kinematics of quasi-elastic lepton-induced single-nucleon knockout from nuclei in the laboratory frame. }
  \label{fig:kinematica}
\end{figure}

The kinematics used in our description of charged-current quasi-elastic neutrino-induced single-nucleon knockout off nuclei is shown in Fig.~\ref{fig:kinematica}. The definitions of the kinematic variables used in the following are the same as in Ref.~\cite{TheMorenoPaper}. An incoming neutrino with four-momentum $k_i^\mu=(E_i,\Vec{k}_i)$ in the xz-plane scatters off a nucleus by exchanging a $W^{\pm}$ boson along the z-axis with four-momentum $Q^\mu=(\omega,\vec{q})=(\omega,0,0,q)$. The four-momentum of the outgoing lepton is denoted by $k_f^\mu = (E_f, \vec{k}_f)$. The momentum transfer $\vec{q}$ is defined as $\vec{q} = \vec{k}_i - \vec{k}_f$. In exclusive CCQE scattering reactions, the interaction with the nucleus results in one detected emitted nucleon with four-momentum 

\begin{align}
    P_N^\mu=(&E_N,\vec{p}_N)\nonumber \\=(&E_N,p_N\sin\theta_N\cos\phi_N, \nonumber \\
    &p_N\sin\theta_N\sin\phi_N,p_N\cos\theta_N), 
\end{align}
with $\theta_N$ the angle of the outgoing nucleon's momentum with the z-axis  and $\phi_N$ the angle between the scattering and reaction plane. The residual nucleus then has a four-momentum 

\begin{align}
    P_{A-1}^\mu=(&E_{A-1},P_{A-1}\sin\theta_{A-1}\cos\phi_{A-1},\nonumber\\&P_{A-1}\sin\theta_{A-1}\sin\phi_{A-1}, P_{A-1}\cos\theta_{A-1})
\end{align}
and is in general in an excited state. In the following, the laboratory frame is used as the frame of reference. In this frame, the initial nuclear complex is at rest, with four-momentum $P_A^\mu=(M_A,\vec{0})$ in natural units, where $M_A$ denotes the rest mass of the nucleus.

In the Born approximation, the cross section can be written as a contraction of the lepton and hadron tensors. The lepton tensor is given by:

\begin{align}
    L_{\mu\nu} &= \overline{\sum_{i,f}} \big(\mathcal{J}_\mu^{lep}(q)\big)^\dagger \mathcal{J}_\nu^{lep} (q),
\end{align}
while the hadron tensor is defined as:
\begin{align}
\label{hadtens}
    W^{\mu\nu}(q) = \overline{\sum_{i,f}} \big(\mathcal{J}^\mu_{had}(q)\big)^\dagger \mathcal{J}^\nu_{had}(q).
\end{align}
In both formulas, $\overline{\sum}_{i,f}$ denotes the sum and average over the relevant leptonic and hadronic quantum numbers, while $\mathcal{J}_\mu(q)$ represents the lepton and hadron currents defined as: 
\begin{align}
\label{currentdef}
    \mathcal{J}_\mu = \langle \Psi_f | \hat{J}_\mu | \Psi_i \rangle,
\end{align}
with $\Psi_i$ and $\Psi_f$ the initial and final state of the system and $\hat{J}_\mu$ the current operator. The current operator for the hadronic vertex in 
charged-current neutrino-nucleus interactions is given in its standard Lorentz-covariant decomposition by~\cite{originalcurrent}:

\begin{align}
    \hat{J}_{had}^\mu 
    &= \Big( F_1(|Q^2|) \gamma^\mu - \frac{F_2(|Q^2|)}{2 m_N} [\gamma^\mu,\gamma^\alpha] Q_\alpha \nonumber\\ &+ G_A(|Q^2|) \gamma^\mu \gamma^5 
    + \frac{G_P(|Q^2|)}{2 m_N} \gamma^5 Q^\mu \Big),
    \label{nucloperator}
\end{align}
with $F_1(|Q^2|)$, $F_2(|Q^2|)$ the Dirac and Pauli vector form factors. $G_A(|Q^2|)$ and $G_P(|Q^2|)$ are the axial and pseudoscalar form factors. Lastly, $\gamma^\mu$ represent the Dirac matrices. The lepton current operator is given by:
\begin{align}
    \hat{J}_\mu^{lep} &= \gamma_\mu(1+h\gamma^5),
\end{align}
which leads to:
\begin{align}
    L_{\mu\nu}=& \frac{2}{m_i m_f}(k_{i,\mu}k_{f,\nu} + k_{f,\nu}k_{i,\mu} - g_{\mu\nu}k_i\cdot k_f+g_{\mu\nu} m_i m_f  \nonumber \\ & - ih\epsilon_{\mu\nu\alpha\beta}k_i^\alpha k_f^\beta).
\end{align}
Here, $g_{\mu\nu}$ represents the metric tensor $(+,-,-,-)$, $\epsilon_{\mu\nu\alpha\beta}$ the Levi-Civita tensor, and $h$ the helicity of the incoming particle. $m_i$ and $m_f$ denote the mass of the incoming and final lepton respectively.

The differential cross section for this reaction can then be written as the contraction of the lepton and hadron tensor as follows:
\begin{align}
    \frac{d^5\sigma}{d\omega d\Omega_f d\Omega_N} = \frac{|\vec{p}_N|M_N}{(2\pi)^3} \sigma^W f_{rec}^{-1} \frac{m_i m_f}{2 E_i E_f}L_{\mu\nu}H^{\mu\nu},
\end{align} 
with  $f_{rec}$ the nuclear recoil factor:
\begin{align}
    &f_{rec} = \Big| 1 + \frac{E_N}{E_{A-1}}\Big(1 - \frac{\vec{p}_N\cdot\vec{q}}{p_N^2}\Big)\Big|,
\end{align}
and $\sigma^W$ a Mott-like prefactor, given by:

\begin{align}
    \sigma^W &= \Big(\frac{G_F \cos\theta_C E_f}{2 \pi}\Big)^2.
\end{align}
Here, $\theta_C$ denotes the Cabibbo angle.

\subsection{Azimuthal asymmetry}

In this section, we sketch how the azimuthal dependence of the hadron tensor, valid for any reaction where one lepton and one hadron is observed, can be derived, building on a general theoretical framework. In such reactions, four four-momenta at the hadronic vertex are relevant for describing the hadron tensor. The incoming momenta given by the momentum transfer $Q^\mu$ and the nuclear target momentum $P_A^\mu$, the outgoing momenta being the outgoing nucleon's momentum $P_N^\mu$ and the residual nucleus' momentum $P_{A-1}^\mu$. These momenta are related to each other by four-momentum conservation: $Q^\mu + P_{A}^\mu = P_N^\mu + P_{A-1}^\mu$. This way, the residual nucleus' recoil momentum can be eliminated so that all invariant structure functions depend on the following six invariants~\cite{TheMorenoPaper}:
\begin{align}
    I_1 &\equiv Q^2 = Q_\mu Q^\mu = \omega^2 - q^2 ,\\
    I_2 &\equiv Q \cdot P_A = Q_\mu P_{A}^\mu = M_A \omega, \\
    I_3 &\equiv Q \cdot P_N = Q_\mu P_N^\mu = E_N \omega - qp_Ncos\theta_N ,\\
    I_4 &\equiv P_A \cdot P_N = P_{A\mu} P_N^\mu = M_A E_N ,\\
    I_5 &\equiv P_N^2 = P_{N\mu} P_N^\mu = M_N^2, \\ 
    I_6 &\equiv P_A^2 = P_{A\mu} P_A^\mu = M_A^2,
\end{align}
where $M_N$ is the mass of the nucleon.
 
The hadron tensor can now be written in terms of the three independent four-momenta $Q^\mu$, $P_N^\mu$ and $P_A^\mu$. It is convenient to replace these four-vectors by the projected four-momenta:

\begin{align}
U^\mu \equiv& \frac{1}{M_A} \left[ P_A^\mu - \left( \frac{Q \cdot P_A}{Q^2} \right) Q^\mu \right] \nonumber\\
=& (1 - \frac{\omega^2}{Q^2},0,0,- \frac{q \omega}{Q^2}),\\
V^\mu \equiv& \frac{1}{M_N} \left[ P_N^\mu - \left( \frac{Q \cdot P_N}{Q^2} \right) Q^\mu \right] \nonumber\\
=& \frac{1}{M_N} (E_N - \frac{E_N\omega^2-q\omega p_N \cos\theta_N}{Q^2}, \nonumber \\ & \> \> p_N \sin \theta_N\cos\phi_N, p_N\sin\theta_N\sin\phi_N, \nonumber \\ & \> \> p_N\cos\theta_N - \frac{E_N\omega q - q^2 p_N \cos\theta_N}{Q^2}), 
\end{align}
\begin{align}
\widetilde{Q}^\mu \equiv& \frac{Q^\mu}{\sqrt{|Q^2|}}\nonumber\\
=& \frac{1}{\sqrt{|Q^2|}}(\omega,0,0,q).
\end{align}

The hadron tensor can be written in terms of symmetric and antisymmetric combinations, denoted by subscripts $s$ and $a$ respectively, of these four-vectors multiplied by functions of the invariants discussed above. Contrary to the decomposition of the hadron tensor used in Refs.~\cite{TheMorenoPaper, DONNELLY1985183}, we also consider higher-order combinations in momentum, comparable to the work of Hernández et al.~for pion production~\cite{hogereorderesponses2}. These are constructed as follows: 

\begin{align}
\label{respeq1}
&{(W^{\mu\nu})^{PC}_s}_{VV/AA} = 
W_{1} g^{\mu \nu} 
+ W_2 U^\mu U^\nu 
+ W_3 \widetilde{Q}^\mu \widetilde{Q}^\nu \nonumber \\ 
&\quad+ W_4 V^{\mu} V^\nu  + W_5 (\widetilde{Q}^\mu U^\nu + U^\mu \widetilde{Q}^\nu ) 
\nonumber\\ &\quad+ W_6 (\widetilde{Q}^\mu V^{\nu} +  V^{\mu} \widetilde{Q}^\nu) 
+ W_7 (U^\mu V^\nu + V^\mu U^\nu ), \\
\nonumber\\
(&{W^{\mu\nu})^{PV}_s}_{VA} = \nonumber \\
&\quad \quad \> W_8 \big(\widetilde{Q}^\mu \epsilon^{\nu}_{\alpha\beta\gamma} V^\alpha U^\beta \widetilde{Q}^\gamma 
+ \widetilde{Q}^\nu \epsilon^{\mu}_{\alpha\beta\gamma} V^\alpha U^\beta \widetilde{Q}^\gamma ) \nonumber \\
&\quad + W_9 \big(U^\mu \epsilon^{\nu}_{\alpha\beta\gamma} V^\alpha U^\beta \widetilde{Q}^\gamma 
+ U^\nu \epsilon^{\mu}_{\alpha\beta\gamma} V^\alpha U^\beta \widetilde{Q}^\gamma ) \nonumber \\
&\quad + W_{10} \big(V^\mu \epsilon^{\nu}_{\alpha\beta\gamma} V^\alpha U^\beta \widetilde{Q}^\gamma 
+ V^\nu \epsilon^{\mu}_{\alpha\beta\gamma} V^\alpha U^\beta \widetilde{Q}^\gamma ), \\
\nonumber
\end{align}
\begin{align}
(&{W^{\mu\nu})^{PV}_a}_{VV/AA} = 
W_{11} (\widetilde{Q}^\mu U^\nu -  U^\mu \widetilde{Q}^\nu) \nonumber \\
&\quad
+ W_{12} (\widetilde{Q}^\mu V^\nu -  V^\mu \widetilde{Q}^\nu)  + W_{13} (U^\mu V^\nu - V^\mu U^\nu), \\
\nonumber\\
(&{W^{\mu\nu})^{PC}_a}_{VA} = 
W_{14} \epsilon^{\mu\nu\alpha\beta} \widetilde{Q}_\alpha U_\beta 
+ W_{15} \epsilon^{\mu\nu\alpha\beta} \widetilde{Q}_\alpha V_\beta \nonumber \\
&\quad + W_{16} \epsilon^{\mu\nu\alpha\beta} U_\alpha V_\beta \nonumber \\
&\quad
+W_{17} \big(\widetilde{Q}^\mu \epsilon^{\nu}_{\alpha\beta\gamma} V^\alpha U^\beta \widetilde{Q}^\gamma 
- \widetilde{Q}^\nu \epsilon^{\mu}_{\alpha\beta\gamma} V^\alpha U^\beta \widetilde{Q}^\gamma ) \nonumber \\
&\quad + W_{18} \big(U^\mu \epsilon^{\nu}_{\alpha\beta\gamma} V^\alpha U^\beta \widetilde{Q}^\gamma 
- U^\nu \epsilon^{\mu}_{\alpha\beta\gamma} V^\alpha U^\beta \widetilde{Q}^\gamma ) \nonumber \\
&\quad + W_{19} \big(V^\mu \epsilon^{\nu}_{\alpha\beta\gamma} V^\alpha U^\beta \widetilde{Q}^\gamma 
- V^\nu \epsilon^{\mu}_{\alpha\beta\gamma} V^\alpha U^\beta \widetilde{Q}^\gamma ).
\label{respeq4}
\end{align}
Here, the coefficients $\textit{W}_{i=1,19}$ represent  functions of the invariants. The subscripts $VV$, $AA$ and $VA$ denote the vector-vector, axial-axial, and vector-axial interference contributions to the hadron tensor. Further, there is a clear distinction between the symmetric and antisymmetric part of the hadron tensor as well as between the nature of the contribution to the tensors: the $PC$ and $PV$ labels correspond to parity conserving and parity violating parts of the hadron tensor~\cite{hogereorderesponses1}. The latter emerge when the relative phase differences of the contributions to the vector and axial current are complex~\cite{KajetanPaperNuwroPhi, SobczykBewijsPV}. These $PC$ and $PV$ terms in the hadron tensor will then also lead to parity conserving and violating contributions to  the cross section. The calculation of the above hadron tensors is shown explicitly in Appendix~\ref{Apprespons}. It has to be noted that higher order terms either lead to the same azimuthal angle distribution or vanish completely because of symmetry considerations\footnote{It is clear that the $\phi_N$-dependence is fully determined by the four-vector $V^\mu$. To end up with different dependencies, higher orders of $V^\mu$ must be considered. The only way to do so without constructing an invariant is placing at least 2 of these four-vectors within the Levi-Civita tensor. Because $\epsilon^\mu_{\alpha\beta\gamma} = - \epsilon^\mu_{\beta\alpha\gamma}$, these contributions will vanish.}. From the azimuthal dependence of the response functions shown in Table~\ref{responssummary}, it is clear that the parity conserving parts of the hadronic responses behave as $\cos\phi_N$ and are hence symmetric for nucleon knockout above and below the interaction plane. The parity violating parts vary as $\sin\phi_N$. This means that the total cross section will behave as $\cos\phi_N$ if the parity violating parts in Eqs.~(\ref{resp2}) and~(\ref{resp3}) vanish, in all other cases the cross section will be a more complicated function of the azimuthal angle containing sines and cosines. Hence, parity violation results in directional asymmetries for the outgoing nucleon. In the following, it will be shown that this fact can be used to distinguish between nuclear interaction models with different degrees of built-in sophistication.

\section{Effect of nucleon distortion}
\label{drie}

The model employed in this work is based on the independent-particle picture of nuclear dynamics, where nucleons move in a  mean-field (MF) potential. It relies on the impulse approximation (IA), which approximates the nuclear many-body current as the incoherent sum of one-body currents acting on individual nucleons~\cite{IA}. In addition, we assume that the initial bound nucleons evolve according to the Schrödinger equation within this MF potential. Their wave functions are obtained through a Hartree-Fock calculation using the Skyrme-type effective nucleon-nucleon interaction SkE2~\cite{skyrme1}, which accurately reproduces ground-state properties and low-lying excitations of nuclei~\cite{skyrme2}. This method has been used extensively in electron-~\cite{RYCKEBUSCH1989694} and neutrino-~\cite{natalieneutrino1, Pandey_2015, Pandey_2016,VanDessel:2017ery,VanDessel:2019atx,VanDessel:2019obk}~nucleus scattering studies, with CRPA corrections on top, and shows a satisfactory agreement with  data.

The used hadronic framework is non-relativistic. The matrix elements are calculated using the non-relativistic projection of the nuclear currents, and the outgoing wave functions are obtained in a non-relativistic manner. Relativistic corrections are taken into account as in Refs.~\cite{non-rela1,non-rela2}. The outgoing wave functions are calculated with a shifted energy transfer to change the outgoing nucleon's momentum to a relativistically consistent value:
\begin{align}
    \omega \to \omega \Big(1+\frac{\omega}{2m_N} \Big),
\end{align}
 meanwhile correcting the final density of states accordingly.
 
For the outgoing nucleon, two treatments are considered here: the plane wave impulse approximation (PWIA) and the distorted wave impulse approximation (DWIA). In the PWIA, final-state interactions (FSI) between the outgoing nucleon and the residual nucleus are neglected, and the ejected nucleon's wave function is described as a plane wave. In the DWIA, the nucleon is modeled as a continuum scattering state in the same MF potential as the initial bound nucleons, ensuring orthogonality with the initial nuclear state. 
 
In both approaches for the description of the outgoing nucleon, the weak charged-current operator is taken as in Eq.~(\ref{nucloperator}). Since all form factors and coefficients in the charged current operator are real, the relative phases between terms will be real as well. Consequently, there is no source of intrinsic parity violation in the PWIA framework. This result is derived analytically in Appendix~\ref{PWIA} and is straightforward to generalize to a fully relativistic framework.\\

In contrast, the DWIA accounts for some final-state interactions between the outgoing nucleon and the residual nucleus. These interactions are modeled through a nuclear potential, which alters the nucleon's wave function and introduces phase shifts. As a consequence, the condition of purely real relative phases, central to the absence of parity-violating effects in the PWIA, is no longer present in the DWIA. The interaction with the nuclear potential can lead to complex amplitudes, thereby enabling parity-violating contributions to the hadron tensor, even when the underlying current operator remains parity conserving. To determine the distorted wave function of the outgoing nucleon in a non-relativistic scheme, the following Schrödinger equation is solved:
\begin{align}
\Big[
\frac{d^2}{dr^2}- &\frac{l(l+1)}{r^2} - \frac{2\eta k}{r}  + k^2\nonumber \\ & - \frac{2m}{\hbar^2}(\hat{V}_C(r)+\hat{V}_{SO}(r)) 
\Big] 
u_{lj}(r) = 0.
\label{schrod}
\end{align}

%
%

In this equation, the second and third terms are the centrifugal barrier and the Coulomb potential, where $l$ is the 
orbital quantum number and $\eta$ the Sommerfeld parameter. $\hat{V}_C(r)$ and $\hat{V}_{SO}(r)$ respectively represent the central- and spin-orbit potential operators. Due to the spin-orbit term, the wave function depends not only on the quantum number $l$, but also on the total angular momentum quantum number $j$. This is represented by the indices of $u_{lj}(r)$. 

\begin{figure}[htbp]
  \centering
  \includegraphics[width=0.48\textwidth]{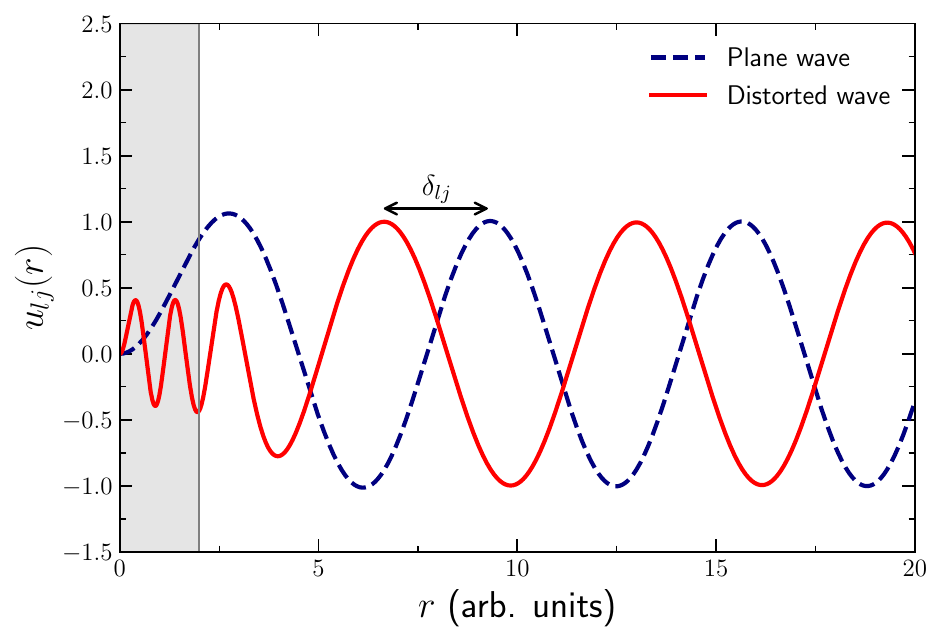} 
  \caption{The effect of the nuclear central potential on the outgoing nucleon's wavefunction. The distorted wavefunction is shifted with a phase $\delta_{lj}$ due to the potential.}
  \label{fig:phaseshift}
\end{figure}

Far away from the nucleus ($r \gg R_A$), the solutions of the Schrödinger equation tend to the Coulomb functions, which behave as sines at large distances~\cite{joachain,Messiah}. With some basic trigonometry, the outgoing nucleon's wave functions can be rewritten for large distances as:
\begin{align}
    \lim_{r \gg R_A} u_{lj}(r) &\sim 
    \sin\Big(kr - \eta \ln(2kr) - l\frac{\pi}{2} + \sigma_{l} + \delta_{lj}\Big) \\
    = &\cos(\delta_{lj}) F_l(\eta,kr) - \sin(\delta_{lj}) G_l(\eta,kr).
\end{align}
Here, $F_l$ and $G_l$ are the regular and irregular Coulomb functions, while $\sigma_l$ and $\delta_{lj}$ are the Coulomb and central phase shifts. They are introduced because the effect of the potential on the outgoing nucleon's wave function is the stretching or squeezing of this wavefunction, compared to the free field wave function. This is visually presented for a Woods-Saxon central potential in Fig.~\ref{fig:phaseshift}, where the Coulomb potential was ignored for the sake of clarity. 

The Coulomb phase shift $\sigma_l$ can be calculated as~\cite{coulombphaseshit}:
\begin{align}
    e^{2i\sigma_l} &= \frac{\Gamma (1 + l + i \eta)}{\Gamma (1 + l - i \eta)},
\end{align}
where $\Gamma$ denotes the gamma function. The phase shift caused by the short-range nuclear potential $\delta_{lj}$ can now be calculated by comparing the numerical solutions of the Schrödinger equation and their derivatives with the expected asymptotic behavior at a large distance $r_m$  from the nucleus. This leads to a simple matrix equation:

\begin{align}
\begin{pmatrix}
u_{lj}(r_{m}) \\
u_{lj}'(r_{m})
\end{pmatrix}
\sim
\begin{pmatrix}
F_l(\eta, kr_{m}) & G_l(\eta, kr_{m})\\
F_l'(\eta, kr_{m}) & G_l'(\eta, kr_{m})
\end{pmatrix}
\begin{pmatrix}
\cos(\delta_{lj}) \\
-\sin(\delta_{lj})
\end{pmatrix}.
\end{align}
Solving this gives rise to the following expression for the central phase shift:
\begin{align}
   \tan^{-1}(\delta_{lj}) &= \frac{G'_l(\eta,kr_{m}) - \frac{u_{lj}'(r_{m})}{u_{lj}(r_{m})}G_l(\eta,kr_{m})}{F'_l(\eta,kr_{m}) - \frac{u_{lj}'(r_{m})}{u_{lj}(r_{m})}F_l(\eta,kr_{m})}. 
   \label{tangens}
\end{align}

\begin{figure}[htbp]
  \centering
  \includegraphics[width=0.48\textwidth]{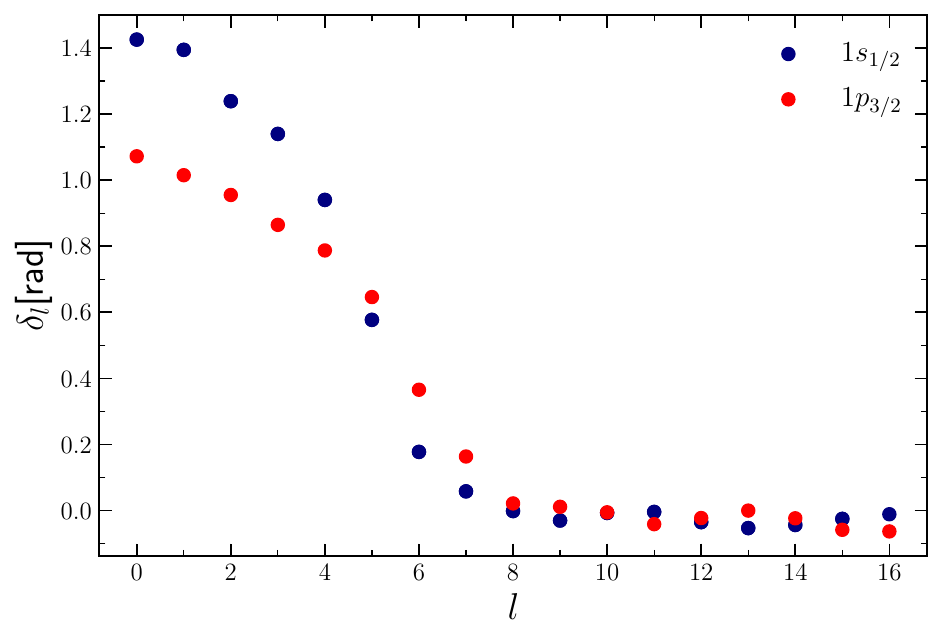} 
  \caption{The central phase shift as a function of quantum number $l$ for a quasi-elastic interaction with $\omega = 80$ MeV. The differences among the shells are caused by the different energy level and quantum number $j$ of both shells.}
  \label{fig:phaseshifts}
\end{figure}

In the Schrödinger equation~(\ref{schrod}),  the relative influence of the nuclear potential is not the same for all combinations of $l$ and $j$. For example, for high orbital momentum states, the effect of the nuclear potential on the phase shift becomes negligible, as can be seen in Fig.~\ref{fig:phaseshifts}. This change in the relative magnitude of the potential affects the central phase shift. Moreover, this quantum number-dependent phase shift will be shown to be the source of the azimuthal asymmetry in the cross section.

The outgoing one-nucleon state can be built in second quantization as a partial wave expansion:

\begin{align}
| \vec{p}_N\,& m_{s_N} \rangle_{PW}
=
\sum_{l,m_l}
\sum_{j,m_j}
(4\pi)\,
\frac{\sqrt{\pi}}{\sqrt{2 M_N \lvert \vec{p}_N \rvert}}\,
i^{\,l}\,
\nonumber\\
&
Y^{*}_{l m_l}(\Omega_N)
\left\langle l, m_l;\,\tfrac{1}{2}, m_{s_N} \,\middle|\, j, m_j \right\rangle\,
\hat{c}^{\dagger}_{l j m_j;PW}\,
\left| \Phi_{\text{g.s.}} \right\rangle ,\\
| \vec{p}_N\,& m_{s_N} \rangle_{DW}
=
\sum_{l,m_l}
\sum_{j,m_j}
(4\pi)\,
\frac{\sqrt{\pi}}{\sqrt{2 M_N \lvert \vec{p}_N \rvert}}\,
i^{\,l}\,
e^{i(\delta_{lj} + \sigma_l)}\,
\nonumber\\
&
Y^{*}_{l m_l}(\Omega_N)
\left\langle l, m_l;\,\tfrac{1}{2}, m_{s_N} \,\middle|\, j, m_j \right\rangle\,
\hat{c}^{\dagger}_{l j m_j;DW}\,
\left| \Phi_{\text{g.s.}} \right\rangle ,
\end{align}
where $\hat{c}^{\dagger}_{l j m_j}$ are the 1p1h creation operators and $\left| \Phi_{\text{g.s.}} \right\rangle$ the nuclear ground state. The $Y_{lm_l} (\Omega_N)$ are spherical harmonics, while $m_l$, $m_j$ and $m_{s_N}$ are the magnetic quantum numbers corresponding to $l$, $j$ and the spin of the outgoing nucleon, respectively. The subscripts $DW$ and $PW$ represent the partial wave expansion for distorted and plane waves. From this equation, it is obvious that the phase shifts for every partial wave will enter the nuclear current as a complex phase. 

\newpage
The azimuthal asymmetry now arises from a combination of two effects:
\begin{itemize}
\item These phases in the nucleon current depend on the combinations of $l$ and $j$ in Eq.~(\ref{tangens}). Due to differing selection rules, the vector and axial contributions to the hadronic tensors are dominated by different $l$ and $j$ contributions. Hence, the phase shifts will not cancel in the VA current contractions. Thus, the combination of the phase shift, the presence of VA terms in PV interactions, and their dependence on the sine of the azimuthal angle, contrary to the cosine behavior of the PC terms (see Table~\ref{responssummary}), results in directional asymmetries for the outgoing nucleon. 
\item Additionally, the $\sin\phi_N$ dependence of the VV and AA terms to $W^{TC'}$ and $W^{TL'}$ will lead to a smaller contribution to the azimuthal asymmetry. These are caused by interferences between  dominant and subdominant contributions to the hadronic tensors.
\end{itemize}

 

It is important to note, though, that the VV and AA parts of the antisymmetric hadronic tensor only enter the cross section formula if the incoming particles are polarized. Hence, contrary to the VA-interferences, this part of the asymmetry is strictly speaking not a pure parity violation effect, but rather a consequence of the polarization of the incoming particles. The AA- and VA-terms only enter the equation in the case of a weakly dominated process. For particles for which the main interaction strength comes from the electromagnetic interaction, these interference terms will be strongly suppressed. 
\\

The $\phi_N$ behavior of the cross section depends on the sophistication of the nuclear model.
As explained in Appendix~\ref{PWIA}, in the PWIA, the x- and y-components of the initial and final nucleon's momentum are the same. Therefore, the $\sin\phi_N$ dependencies in the responses cannot arise. This means that the azimuthal angle distribution will be symmetric around $\pi$. However, in the DWIA, the interferences will cause differences between the x- and y-components of the initial and final nucleon's momentum. This causes $\sin\phi_N$ and $\sin 2\phi_N$ dependencies in the hadronic responses, breaking this symmetry. To conclude this section, differences between theoretical models are not the only piece of information that the azimuthal angle distribution delivers. In the nuclear shell model, all shells have different combinations of quantum numbers $l$ and $j$, so each shell will result in an outgoing nucleon with distinct $l$ and $j$ according to the selection rules. The Coulomb and central phase shifts of these outgoing nucleons will thus be different for nucleons originating from different shells and will lead to different $\phi_N$ distributions of the cross sections for nucleons coming from each shell. The azimuthal angle thus also carries information on the shell structure of the parent nucleus. In the remaining sections, the applications and feasibility of detecting this azimuthal asymmetry will be studied.

%% file: Sections/Results.tex
\section{Model validation and results}
\label{vier}

\begin{figure*}[htbp]
\includegraphics[width=1\linewidth]{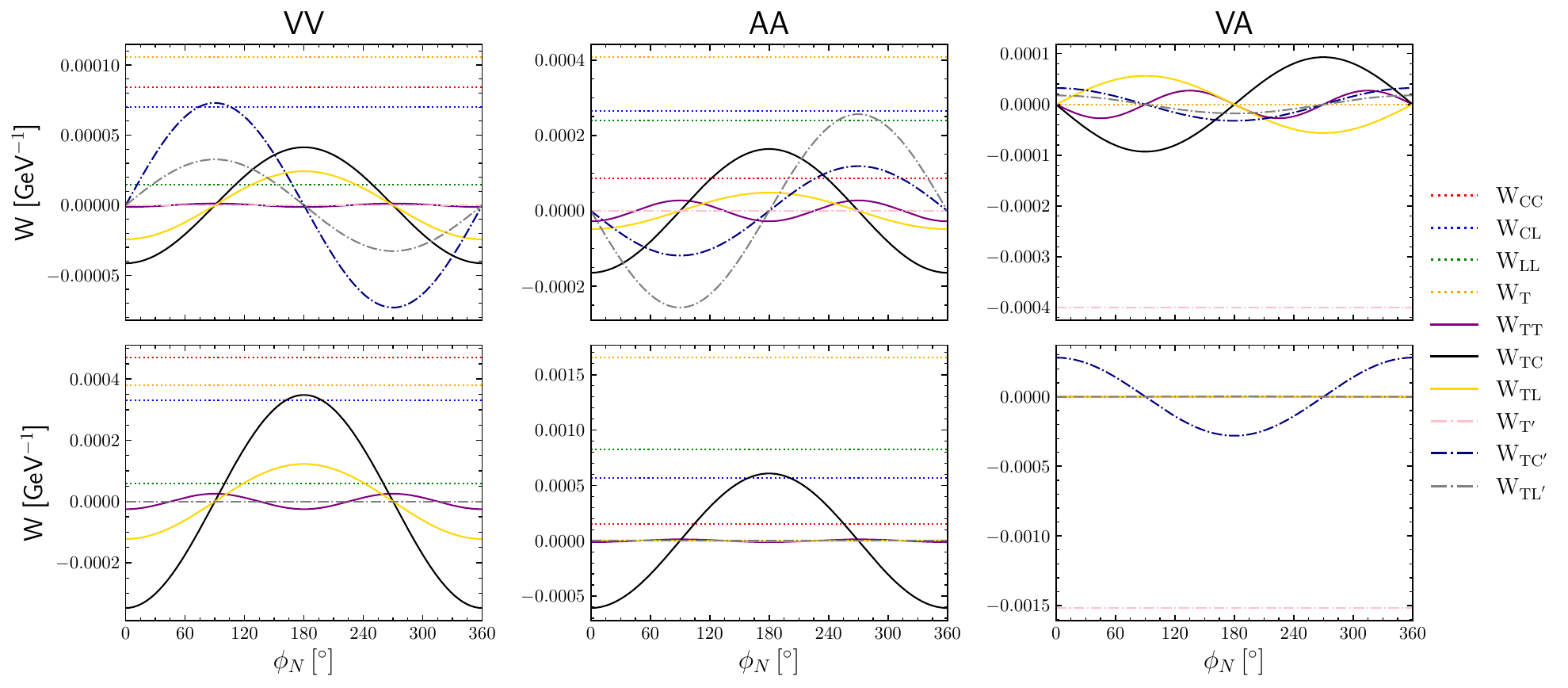}
  
  \caption{Azimuthal angle distributions of the hadronic response functions for quasi-elastic
$\nu_\mu + {}^{12}\!\mathrm{C} \to \mu^- + {}^{11}\!\mathrm{C} + p^+$ interactions with an energy transfer
$\omega = 160$~MeV, outgoing lepton and nucleon polar angles fixed at 
$\theta_l = \theta_N = 30^\circ$, and an incoming neutrino energy
$E_\nu = 400$~MeV. The top row shows the DWIA results, while the bottom row shows the PWIA results.}

  \label{respo2}
\end{figure*}

The numerical results of the models are validated against the theoretically expected behavior. In Fig.~\ref{respo2}, the azimuthal angle distributions of all response functions are shown for the CCQE reaction with a muon-neutrino energy $E_\nu = 400$ MeV on $^{12}$C, with $\omega = 160$ MeV and $\theta_l = \theta_N = 30\degree$. Here, $\theta_l$ denotes the outgoing lepton's polar angle. For both the PWIA and the DWIA, the distributions follow the behavior predicted in Appendix~\ref{Apprespons}. For the PWIA results, no sinusoidal dependencies are observed, indicating that the outgoing nucleon has no preferred direction out of the scattering plane. In contrast, the DWIA calculations exhibit such sinusoidal dependencies, leading to a preferred emission direction of the nucleon out of the scattering plane. As discussed in earlier sections, this effect mainly originates from parity violation.

\begin{figure*}[htbp]
    \includegraphics[width=1\linewidth]{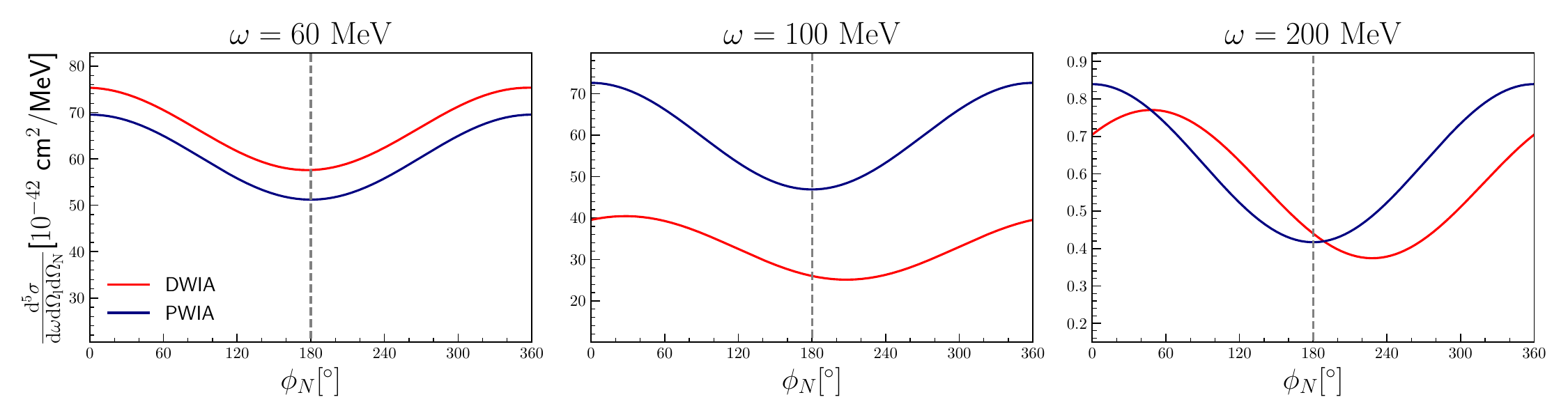}
  
  \caption{Azimuthal angle distributions of the differential cross section for quasi-elastic $\nu_\mu + {}^{12}\!\mathrm{C} \to \mu^- + {}^{11}\!\mathrm{C} + p^+$ interactions with outgoing lepton and nucleon polar angles fixed at 
$\theta_l = \theta_N = 30^\circ$ for different energy transfers. The incoming neutrino energy is $E_\nu = 700$~MeV. The gray dashed line marks $\phi_N = 180^\circ$ as a reference.}
\label{fig:E700C12}
\end{figure*}

\begin{figure*}[htbp]
    \includegraphics[width=1\linewidth]{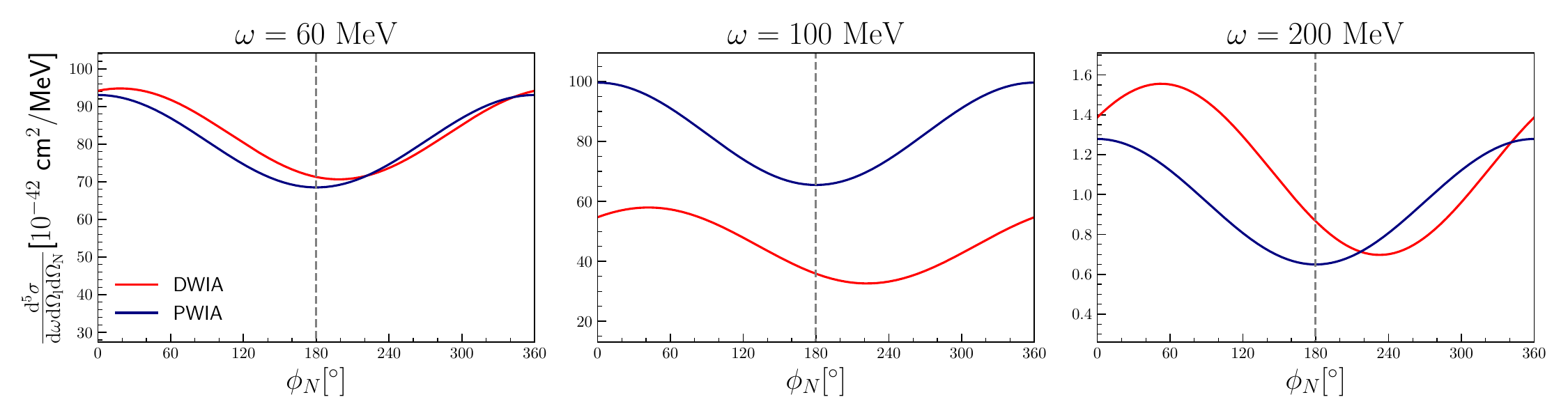}
  \caption{Azimuthal angle distributions of the differential cross section for quasi-elastic
$\nu_\mu + {}^{16}\!\mathrm{O} \to \mu^- + {}^{15}\!\mathrm{O} + p^+$ interactions with outgoing lepton and nucleon polar angles
fixed at $\theta_l = \theta_N = 30^\circ$ for different energy transfers. The incoming neutrino energy is $E_\nu = 700$~MeV. The gray dashed line marks $\phi_N = 180^\circ$ as a reference.}
\label{fig:E700O16}
\end{figure*}

The effect of the parity violating contributions on the differential cross section as a function of the azimuthal angle can be seen in Fig.~\ref{fig:E700C12} for CCQE reactions on $^{12}$C and in Fig.~\ref{fig:E700O16} for reactions on $^{16}$O. For both figures, the incoming muon-neutrino energy is $E_\nu = 700$ MeV and the outgoing lepton and nucleon polar angles are $\theta_l = \theta_N=30 ^\circ$. It is observed that in both cases the results calculated using the PWIA are symmetric around $\phi_N = 180^\circ$, while for the DWIA calculations, directional asymmetries are present. 

\begin{figure*}[htbp]
   \includegraphics[width=1\linewidth]{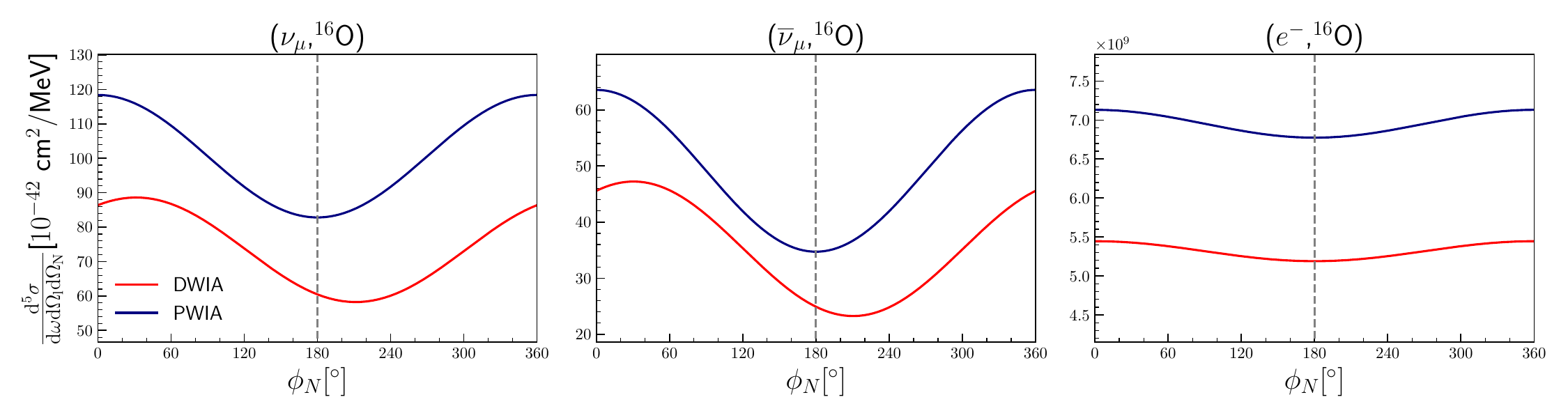}  
  \caption{Azimuthal angle distributions of the differential cross section for charged-current quasi-elastic scattering. Panels show neutrino- (left), antineutrino- (middle), and electron- (right) nucleus interactions, with outgoing lepton and nucleon polar angles fixed at $\theta_l = \theta_N = 30^\circ$. The incoming neutrino energy is $E_\nu = 700$~MeV and the energy transfer is $\omega = 80$~MeV. The gray dashed line marks $\phi_N = 180^\circ$ as a reference.}
\label{fig:pvE700O16}
\end{figure*}

The parity-violating nature of this effect is illustrated in Fig.~\ref{fig:pvE700O16}, where the differential cross section is shown for incoming muon neutrino, muon antineutrino, and electron scattering off $^{16}$O. The energy of the incoming particle is 700~MeV. The outgoing lepton and nucleon polar angles are $\theta_l = \theta_N=30^\circ$, and the energy transfer is $\omega = 80$~MeV. Both neutrino and antineutrino cross sections exhibit sizable directional asymmetries, which are absent in electron–nucleus scattering. As in electron-nucleus scattering the cross section is dominated by the electromagnetic interaction, the sine dependencies corresponding to the PV VA-interference terms originating from the weak interaction are not present. Additionally, if the electron beam is unpolarized, the remaining sine dependencies vanish. If the directional asymmetries were solely due to the polarization of the incoming particle, the $\phi_N$ distributions for neutrino and antineutrino scattering would be mirror symmetric around $\phi_N = 180^\circ$ because of the opposite-signed helicity. As this is not the case, the parity-violating nature of the weak interaction plays a significant role in shaping the asymmetries in the azimuthal distribution of the outgoing nucleon.

\begin{figure*}[htbp]
\includegraphics[width=1\linewidth]{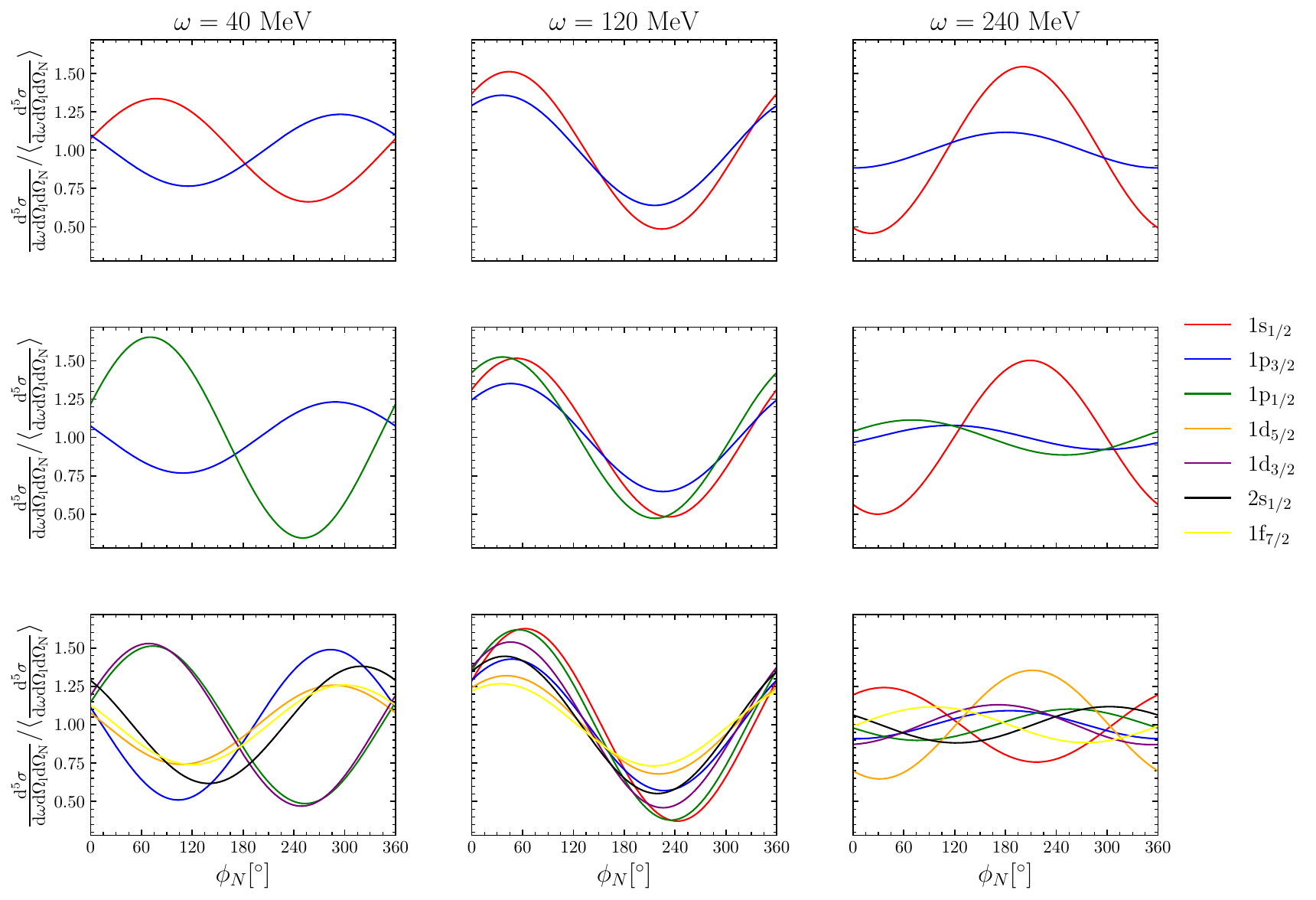}
	\caption{Normalized differential cross sections for nucleons from different shells for $\nu_\mu$-nucleus interactions on $^{12}$C (top), $^{16}$O (middle) and $^{40}$Ar (bottom) for energy transfers $\omega= 40, 120, 240$ MeV. The incoming neutrino energy is $E_\nu = 400$ MeV, the muon and nucleon polar angles are fixed at $\theta_l=\theta_N=30\degree$. The binding energies of the $1s_{1/2}$ shells in $^{16}$O and $^{40}$Ar exceed 40 MeV, thus a boson with $\omega = 40$ MeV can not emit a nucleon coming from that shell.}
	\label{shells}
\end{figure*}

As discussed in section~\ref{twee}, the $\phi_N$ distribution is expected to differ for nucleons originating from different nuclear shells. In Fig.~\ref{shells}, the azimuthal angle distributions of the differential cross sections for $\nu_\mu$-interactions on $^{12}$C, $^{16}$O and $^{40}$Ar are shown separated into their distinct shell contributions. Each shell contribution is normalized independently to allow for a comparison of the distribution shapes. One can see that nucleons originating from shells with different quantum numbers exhibit distinct $\phi_N$ distributions, indicating that measurements of the azimuthal-angle distribution in neutrino–nucleus scattering can indeed provide information on the nuclear shell structure. The $\phi_N$ distribution also depends on the energy transfer $\omega$. Since the binding energies vary among shells, the emitted nucleons acquire different kinetic energies and corresponding wavelengths. As a result, the phase shifts affect nucleons from different shells differently, even for identical boson kinematics. Consequently, the azimuthal-angle distributions of nucleons from equivalent shells also vary between nuclei. Although such shells share the same quantum numbers, their binding energies depend on the nucleus, reflecting differences in the underlying nuclear mean-field potential.

%% file: Sections/Experiment.tex
\section{Experimental sensitivity and prospects for long-baseline measurements}
\label{vierb}

In this section, we translate the theoretically obtained $\phi_N$ asymmetry into an experimentally accessible observable, quantify its robustness against energy-reconstruction biases and inelastic FSI, and discuss the prospects for observing it with accelerator neutrinos. The definition of $\phi_N$ used in the previous sections is the azimuth of the outgoing hadron around the momentum transfer $\vec{q}$ with respect to the lepton scattering plane. Concretely, this can be reconstructed experimentally with the following considerations:
\begin{itemize}
\item the beam direction provides a good approximation for the incoming neutrino direction in the case of long-baseline experiments (such as the Tokai-to-Kamiokande (T2K) experiment);
\item the beam and muon directions define the lepton reaction plane;
\item the momentum transfer $\vec{q}\equiv \vec{k}_i-\vec{k}_f$, contained in the lepton reaction plane, defines the longitudinal axis, where $\vec{k}_i$ and $\vec{k}_f$ denote the reconstructed incoming neutrino and outgoing muon momenta, respectively;
\item $\phi_N$ is the nucleon azimuth around $\vec{q}$, after rotating into the interaction frame and choosing the lepton plane such that the muon azimuth is set to zero.
\end{itemize}

In most accelerator neutrino experiments, $\vec{k}_i$ is not known on an event-by-event basis because $E_\nu$ is unknown. Two practical strategies are commonly used:
    (i) Quasi-elastic kinematic reconstruction (using only the muon, assuming a fixed removal energy), or
    (ii) Combining muon kinematics with the reconstructed hadronic energy, also assuming a specific removal energy.

Both approaches introduce model dependence through the assumed removal (or binding) energy. This is precisely where the nuclear shell structure becomes relevant: deeper shells imply larger removal energies, which biases the inferred $\vec{q}$ and, consequently, the reconstructed $\phi_N^{reco}$.

In the case of a charged-current quasi-elastic interaction on a nuclear target with an observable hadronic energy, the neutrino energy can be expressed as:
\begin{equation}
E_\nu^{\rm cal} \simeq E_\mu + E_{\rm N}^{\rm vis} + \bar{E}_b,
\label{eq:enu_cal_simple}
\end{equation}
where $E_{\rm N}^{\rm vis}$ denotes the visible energy carried by reconstructed final-state nucleons (typically the sum of their reconstructed kinetic energies above threshold), and $\bar{E}_b$ represents the average removal (or binding) energy.

Because the struck nucleon is emitted from within the nuclear medium, inelastic final-state interactions (FSI) can result in final states containing more than one outgoing nucleon. Two strategies may be employed depending on the detector technology: (i) selecting only the leading hadron emerging from the nucleus, or (ii) summing the energies of all emitted particles, as done for the MINER$\nu$A visible energy observable~\cite{intro4}. We adopt the first approach, which aligns more closely with the T2K-like methodology, where $E_{\rm N}^{\rm vis}$ corresponds to the energy of the leading nucleon in the final state when constructing $E_\nu^{\rm cal}$.

For T2K and, in particular, for the upgraded near detector featuring SuperFGD~\cite{Blondel:2020hml}, a representative detection threshold for protons is $p_{\rm thr} \simeq 300~\mathrm{MeV}/c$~\cite{Dolan:2021hbw}. In the following, we adopt this value as a benchmark for proton detectability, while noting that future experiments may achieve improved acceptance and lower thresholds. A fully realistic T2K-specific analysis would require incorporating angular and energy resolutions (as well as efficiencies) as functions of particle momentum; however, such detector-specific effects are beyond the scope of the present study.

Finally, since the initial shell cannot be identified event-by-event, the binding energy must be approximated. Here we adopt a fixed average value $\bar{E_b} = 28~\mathrm{MeV}$ for a target $^{12}\mathrm{C}$, which corresponds to the weighted average of the binding energies of the two shells by their respective total cross-sectional values for the studied neutrino energy spectrum.

For each shell, $1s_{1/2}$ and $1p_{3/2}$, a dataset of one million exclusive $1p1h$ events was generated in the impulse approximation using the Normalizing-Flow-based methodology~\cite{Elbaz2025}. The neutrino energies were sampled from the T2K flux truncated at $1.3$~GeV, which covers the main flux support relevant for charged-current quasielastic event production while suppressing the high-energy tail, where multinucleon and pion-production channels become dominant. In addition, a combined-shell sample of one million events was constructed by mixing the $1s_{1/2}$ and $1p_{3/2}$ contributions according to their relative partial cross sections, approximately following the expected shell-occupancy ratio of $1:2$. To account for inelastic final-state interactions, the NEUT intranuclear cascade model~\cite{Hayato:2009zz} was applied to all events.

The proton azimuthal angle $\phi_N^{reco}$ was reconstructed according to the procedure described above, using $p_{\rm thr} \simeq 300~\mathrm{MeV}/c$ as the proton detection threshold for defining the visible nucleon energy. Specifically, we set the incident neutrino momentum along the beam direction, $\vec{k}_i = E_\nu^{\rm cal}\,\hat{n}_{\rm beam}$ (where $\hat{n}_{\rm beam}$ is the beam unit vector), and compute the momentum transfer $\vec{q} = \vec{k}_i - \vec{k}_f$. We then rotate to a frame where $\hat{z} \parallel \vec{q}$ and choose $\hat{x}$ within the lepton-scattering plane such that $\phi_\mu = 0^\circ$; consequently, $\phi_N^{reco}$ represents the azimuth of the outgoing nucleon around $\vec{q}$. In this study, we restricted the analysis to the leading reconstructed proton (i.e., the one with the largest visible energy) and discarded events in which no proton passed the detection threshold.

\begin{figure}[h!]
  \includegraphics[width=0.49\textwidth]{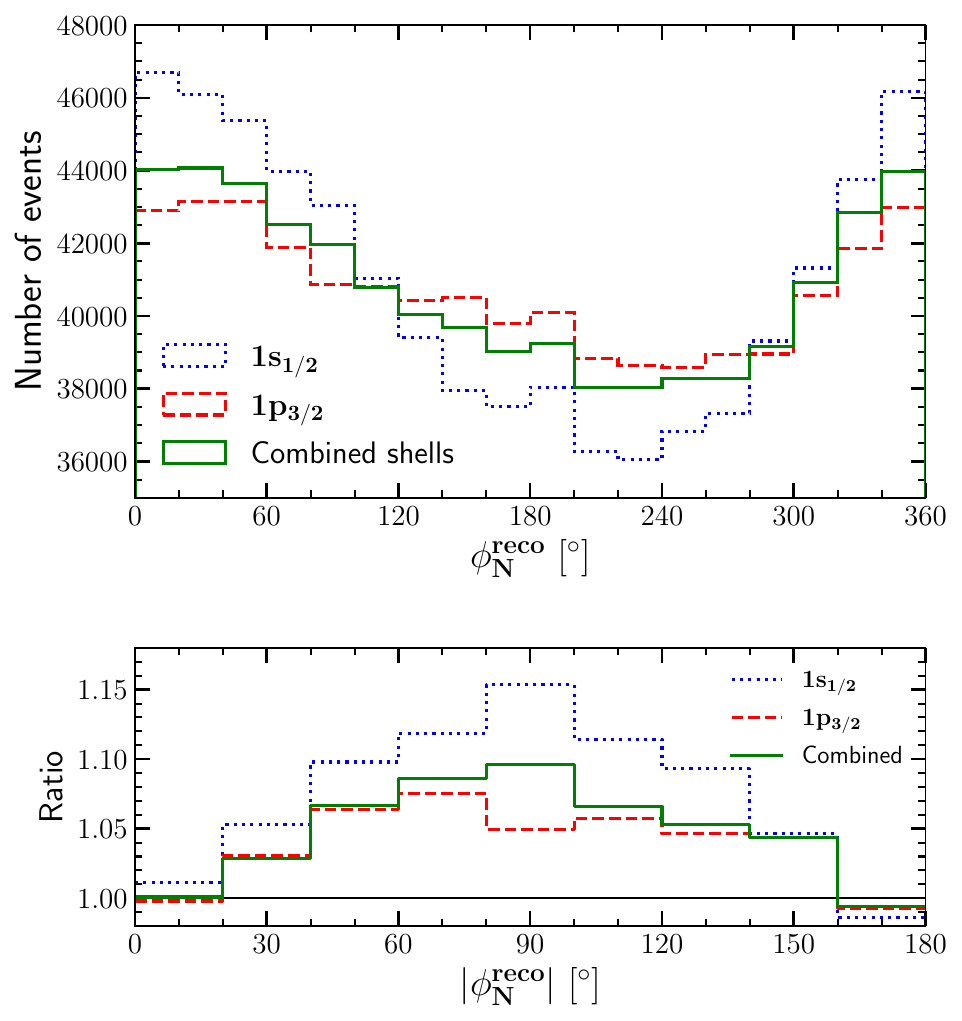}
  \caption{Top: Distribution of the reconstructed leading-proton azimuthal angle $\phi_N^{\rm reco}$ for the $1s_{1/2}$ and $1p_{3/2}$ shells, together with the combined-shell sample, after applying the NEUT intranuclear cascade model. Bottom: ratio between the above-plane and below-plane hemispheres, computed as $N(\phi_N^{\rm reco})/N(360^\circ-\phi_N^{\rm reco})$ over the range $\phi_N^{\rm reco}\in[0^\circ,180^\circ]$, shown separately for each shell and for the combined-shell sample.}
  \label{fig: phi distribution shell}
\end{figure}

Fig.~\ref{fig: phi distribution shell} shows the distributions of the reconstructed leading-proton azimuthal angle $\phi_N^{\rm reco}$ for the $1s_{1/2}$ and $1p_{3/2}$ shells of the $^{12}\mathrm{C}$ nucleus, as well as for the combined-shell sample obtained by weighting the two contributions according to their respective partial cross sections. After applying the NEUT intranuclear cascade model, the two shells exhibit similar but distinct angular behaviors. The $1s_{1/2}$ shell shows the most pronounced asymmetry between the two hemispheres, with an excess of leading protons emitted above the leptonic reaction plane, corresponding to $\phi_N^{\rm reco}\in[0^\circ,180^\circ]$. The $1p_{3/2}$ shell is more uniform and closer to symmetric, although it still shows a smaller excess in the same hemisphere. The combined-shell sample retains this effect, with an asymmetry intermediate between the two individual shell contributions.

The shell dependence therefore induces an asymmetric distribution of emitted protons above and below the leptonic reaction plane. To quantify this effect, we define the up--down asymmetry observable
\begin{equation}
A \equiv \frac{N_{\rm above} - N_{\rm below}}{N_{\rm above} + N_{\rm below}},
\label{eq:aplanarity_def}
\end{equation}
where $N_{\rm above}$ and $N_{\rm below}$ denote the number of events in which the leading reconstructed proton is emitted above and below the leptonic reaction plane, respectively. In practice, after mapping $\phi_N^{\rm reco}$ to the range $[0^\circ,360^\circ]$, the above-plane hemisphere is defined by $0^\circ < \phi_N^{\rm reco} < 180^\circ$, while the below-plane hemisphere is defined by $180^\circ < \phi_N^{\rm reco} < 360^\circ$.

The asymmetry $A$ is evaluated independently for each shell as a function of the generated sample size, with the results reported in Table~\ref{tab:aplanarity_bootstrap}. The quoted uncertainties correspond to the $68\%$, $95\%$, and $99\%$ confidence intervals obtained from $5000$ bootstrap resamplings of the full simulated sample for each shell. We also report the corresponding observable for the combined-shell sample, where the $1s_{1/2}$ and $1p_{3/2}$ contributions are mixed according to their relative cross-section weights.

The results show that a generated sample of $5\times 10^{3}$ events is already sufficient to observe a non-zero asymmetry in the combined-shell sample at the $68\%$ confidence level. The effect remains statistically separated from zero at the $95\%$ level for samples of $10^{4}$ events, and at the $99\%$ level for samples of $1.5\times 10^{4}$ events. This indicates that the shell-dependent up--down asymmetry survives inelastic final-state rescattering and realistic neutrino energy reconstruction, making it a promising observable for near-detector measurements. Translating these requirements into specific event counts for T2K will depend on detector efficiencies, flux profiles, and background contamination in this event topology. Nevertheless, the required statistics are modest, suggesting that such a measurement could be feasible in a T2K-like detector with good tracking performance.

\begin{table*}[t]
\centering
\caption{Up--down asymmetry $A$ (in percent) for the $1s_{1/2}$ and $1p_{3/2}$ shells, and for both shells combined weighted their relative contributions, as a function of the generated Monte Carlo sample size. The central value is quoted together with asymmetric confidence intervals: the first uncertainty corresponds to the $68\%$ confidence interval, the values in the first parentheses correspond to the $95\%$ confidence interval, and the values in the second parentheses correspond to the $99\%$ confidence interval. Intervals are obtained from $5000$ bootstrap resamplings. The observable is computed using the leading reconstructed proton in each event, and events with no proton above the detection threshold are excluded.}
\begin{tabular}{l@{\hspace{0.5cm}}c@{\hspace{0.5cm}}c@{\hspace{0.5cm}}c}
\toprule
\textbf{Number of protons} &
\textbf{$1s_{1/2}$: $A$ [\%]} &
\textbf{$1p_{3/2}$: $A$ [\%]} &
\textbf{Combined: $A$ [\%]} \\
\midrule
$5000$ &
$3.52^{+1.44}_{-1.40}\ (^{+2.88}_{-2.72})\ (^{+3.72}_{-3.52})$ &
$1.93^{+1.39}_{-1.41}\ (^{+2.83}_{-2.65})\ (^{+3.63}_{-3.61})$ &
$2.30^{+1.38}_{-1.38}\ (^{+2.78}_{-2.78})\ (^{+3.62}_{-3.62})$ \\
$10000$ &
$3.52^{+0.98}_{-1.00}\ (^{+1.92}_{-1.96})\ (^{+2.46}_{-2.56})$ &
$1.93^{+1.03}_{-0.97}\ (^{+1.95}_{-1.91})\ (^{+2.51}_{-2.67})$ &
$2.32^{+1.00}_{-1.02}\ (^{+2.02}_{-2.00})\ (^{+2.60}_{-2.64})$ \\
$15000$ &
$3.55^{+0.82}_{-0.82}\ (^{+1.66}_{-1.58})\ (^{+2.13}_{-2.09})$ &
$1.91^{+0.81}_{-0.82}\ (^{+1.58}_{-1.66})\ (^{+2.01}_{-2.17})$ &
$2.32^{+0.82}_{-0.82}\ (^{+1.65}_{-1.63})\ (^{+2.04}_{-2.20})$ \\
$20000$ &
$3.54^{+0.70}_{-0.70}\ (^{+1.39}_{-1.39})\ (^{+1.88}_{-1.80})$ &
$1.92^{+0.71}_{-0.72}\ (^{+1.36}_{-1.39})\ (^{+1.82}_{-1.77})$ &
$2.31^{+0.70}_{-0.71}\ (^{+1.37}_{-1.40})\ (^{+1.77}_{-1.87})$ \\
\bottomrule
\end{tabular}
\label{tab:aplanarity_bootstrap}
\end{table*}

%% file: Sections/Conclusion.tex
\section{Conclusion}
\label{vijf}

In this work, the general formalism for exclusive charged-current interactions from Refs.~\cite{TheMorenoPaper, DONNELLY1985183} is extended to include higher order terms in momenta. Using this framework, it was demonstrated that in the case of charged-current quasi-elastic neutrino-nucleus scattering, the azimuthal dependence of the outgoing nucleon shows non-trivial directional asymmetries. The cause of this phenomenon was identified to be originating from the parity-violating nature of the weak interaction.

It was demonstrated that the azimuthal angle distribution of the outgoing nucleon is sensitive to the nuclear interaction model employed in the description. More specifically, it was shown that plane wave impulse approximation approaches do not introduce such directional asymmetries, while a consistent description of the outgoing wave as a distorted wave does lead to said asymmetries. The validity of a plane wave description of the final nucleon for experimental exclusive studies can hence be expected to be limited. The angular distribution of the outgoing nucleons also depends on the nuclear shell from which the nucleon originated. This is caused by the dependence of the phase shift on orbital and total angular momentum quantum numbers $l$ and $j$, which are different for  specific nuclear shells.

The experimental feasibility of observing this effect in current long-baseline neutrino oscillation experiments was studied. To assess this, the NEUT intranuclear cascade model was used to account for final-state interactions. Additionally, a realistic momentum detection threshold was applied. It was found that the azimuthal asymmetries survived the final-state interactions and nucleon-threshold selection. The asymmetry was estimated to be visible with $\mathcal{O}$($10^4$) events at the $99\%$ confidence level, suggesting that the effect is visible in a T2K-like detector.